\documentclass[aps,prb,twocolumn,superscriptaddress]{revtex4-2}
\usepackage{amsmath,amssymb}
\usepackage{graphicx}
\usepackage{bm}

\begin{document}

\title{Symmetry-protected intraband exciton-phonon scattering and its breakdown by mass asymmetry}

\author{Michael O.~Atambo}
\affiliation{Department of Physics Earth and Environmental Science, Technical University of Kenya, Nairobi, Kenya}

\begin{abstract}
The coupling of excitons to lattice vibrations is typically treated via phenomenological models that artificially separate long-range Fr\"ohlich and short-range Holstein interactions. Recent analytical work has established that, in the delocalized limit, Fr\"ohlich scattering is suppressed by electron--hole interference. Here we present a model-space Bethe--Salpeter framework to evaluate exciton--phonon coupling across the extended-to-localized crossover. We establish a fundamental distinction between the \emph{inclusive} exciton--phonon coupling weight, obtainable via an exact completeness sum rule without summation over excited states, and the \emph{exclusive} intraband (internal-state-preserving) scattering amplitude that governs low-energy decoherence. We prove an exact symmetry theorem: for an inversion-symmetric relative-coordinate Hamiltonian with equal electron and hole masses, the intraband Fr\"ohlich vertex vanishes identically, protecting the exciton from low-energy polar phonon scattering. When mass asymmetry is introduced, the finite-momentum relative wavefunction acquires a complex phase twist that breaks this protection. By analyzing the long-wavelength limit, we derive a controlled small-$q$ activation law showing that the intraband vertex scales as $F_{00}(q) \propto \Delta q^2 \langle r^2 \rangle$, where $\Delta$ parameterizes the mass asymmetry. Finally, we compute the second-order polaron self-energy shift and demonstrate that mass asymmetry dramatically enhances phonon dressing, confirming that the symmetry theorem directly governs the many-body energy renormalization of the exciton. These results provide a rigorous conceptual framework for understanding the competition between long-range and local exciton--phonon coupling in polar semiconductors.
\end{abstract}

\maketitle

\section{Introduction}

The accurate prediction of excitonic properties in semiconductors and
insulators has been revolutionized by many-body perturbation theory
(MBPT), particularly through the combination of $GW$ quasiparticle
corrections~\cite{Hedin1965,Hybertsen1986} and the Bethe-Salpeter
equation (BSE) for optical spectra~\cite{Strinati1980,Strinati1982,
Onida2002}. These frameworks have achieved quantitative agreement with
experiment for a wide range of materials, from bulk semiconductors to
wide-gap transition-metal oxides~\cite{Atambo2019,Sangalli2019}.
However, a persistent challenge remains: the coupling of excitons to
lattice vibrations (phonons), which governs critical phenomena including
exciton linewidths, phonon sidebands, temperature-dependent binding
energies, and polaron formation~\cite{Giustino2017}.

In practice, exciton–phonon interactions are often rationalized in terms of two limiting pictures. The \emph{Fr\"ohlich
model}~\cite{Frohlich1954} describes the interaction of charge carriers
with the macroscopic electric field of longitudinal optical (LO) phonons
and is appropriate for delocalized Wannier-Mott
excitons~\cite{Wannier1937,Mott1938} in polar materials. The
\emph{Holstein model}~\cite{Holstein1959}, by contrast, describes local
deformation-potential coupling and is appropriate for tightly bound
Frenkel excitons~\cite{Frenkel1931} in molecular crystals and strongly
correlated insulators. The literature typically treats these as mutually
exclusive regimes, with the choice of model guided by qualitative
arguments about the exciton Bohr radius relative to the lattice
constant~\cite{Toyozawa2003,Emin1973}.

This dichotomy is unsatisfactory for several reasons. First, many
materials of current interest-halide perovskites, transition-metal
dichalcogenides (TMDs), and transition-metal oxides-host excitons
that are intermediate between the Wannier and Frenkel limits, with
localization lengths comparable to a few lattice
spacings~\cite{Chernikov2014,Berkelbach2013,Atambo2019}. Second, the
Fr\"ohlich model assumes a delocalized hydrogenic envelope and a
well-defined dipole moment; when the exciton localizes, the multipole
structure of the coupling changes fundamentally, and the Fr\"ohlich
approximation can fail catastrophically~\cite{Toyozawa1956,Rashba1957}.
Third, and most critically, recent analytical work has shown that even
in the Wannier limit, the elastic Fr\"ohlich coupling between excitonic
states of the same parity is exactly suppressed by destructive
electron-hole interference~\cite{Atambo2024parity}. This parity
selection rule implies that the ground-state exciton is protected from
long-wavelength polar phonons, a result that has profound implications
for exciton transport and coherence but whose domain of validity remains
unexplored.

The central question addressed in this work is therefore:
\emph{Under what conditions does the parity selection rule break down,
and how does the exciton-phonon coupling evolve continuously from the
Wannier to the Frenkel limit?}

Answering this question from first principles is computationally
prohibitive. A full \emph{ab initio} treatment requires solving the
BSE~\cite{Strinati1980,Strinati1982,Rohlfing2000,Albrecht1998} on
phonon-distorted lattices, computing electron-phonon matrix elements
via density-functional perturbation theory (DFPT), and summing over all
intermediate excitonic states-a procedure that scales poorly with
system size and is currently limited to small unit
cells~\cite{Antonius2021,Hartmann2022}. This computational bottleneck
has prevented a systematic exploration of the Wannier-Frenkel crossover
and the mass-asymmetry dependence of exciton-phonon coupling.

Here we overcome this limitation by developing a \emph{model-space}
methodology that is both numerically exact within its Hilbert space and
computationally trivial. Our approach rests on three key insights:

\begin{enumerate}
\item \textbf{Exact sum rules.} The total exciton-phonon coupling
weight, summed over all final states, can be evaluated as a ground-state
expectation value $\langle 0|\hat{O}^\dagger \hat{O}|0\rangle$ via the
completeness relation. This eliminates the need to explicitly sum over $\mathcal{O}(L)$ excited states per momentum transfer, where $L$ is the number of lattice sites.

\item \textbf{Gauge-consistent vertices.} We construct the exciton-phonon
vertex operators in the same center-of-mass gauge as the model-space BSE
Hamiltonian, ensuring that the parity selection rule emerges naturally
from the symmetry of the problem rather than being imposed by hand.

\item \textbf{Analytical scaling laws.} In the continuum limit, we derive
closed-form expressions for the elastic Fr\"ohlich and Holstein vertices
as functions of the mass fractions $\alpha_e$, $\alpha_h$ and the
localization length $\xi$. These yield a robust scaling criterion $W_H/W_F \propto [\Delta^2 (\xi/a)^2]^{-2}$ within the model space.
\end{enumerate}

Our model Hamiltonian is a one-dimensional tight-binding BSE with
tunable electron-hole attraction, which allows us to continuously
interpolate between the Wannier limit ($\xi \gg a$, where $a$ is the
lattice constant) and the Frenkel limit ($\xi \sim a$). Despite its
simplicity, this model captures the essential physics of the crossover:
the competition between long-range Coulomb screening, kinetic
delocalization, and local lattice deformation. We validate the sum-rule approach by explicit summation over intermediate excitonic states and compute the second-order perturbative self-energy to connect the matrix-element theorems to many-body energy renormalization.

The principal results of this work are:
\begin{itemize}
\item The parity selection rule for elastic Fr\"ohlich scattering is exact within the inversion-symmetric equal-mass model, independent of the exciton localization length.

\item Mass asymmetry ($m_e \neq m_h$) breaks the parity protection by inducing a complex phase twist in the moving exciton wavefunction, activating a low-$q$ intraband scattering channel that scales as $F_{00}(q) \propto \Delta q^2 \langle r^2 \rangle$.
\item As the exciton localizes ($\xi \to a$), the long-range Fr\"ohlich channel is geometrically starved relative to the local Holstein channel, driving a crossover in the dominant coupling mechanism.
\item The inclusive (Franck-Condon) coupling weight remains finite in the Frenkel limit due to interband transitions, while the exclusive intraband (decoherence) weight is strictly governed by mass asymmetry and localization.

\end{itemize}

The remainder of this paper is organized as follows. In
Sec.~\ref{sec:model}, we define the model-space BSE Hamiltonian and the
exact sum-rule methodology. In Sec.~\ref{sec:analytical}, we derive the
continuum-limit scaling laws and the master crossover equation. In
Sec.~\ref{sec:results}, we present numerical results across the
Wannier-Frenkel crossover and compare with the analytical predictions.
We discuss implications for real materials in Sec.~\ref{sec:discussion}
and conclude in Sec.~\ref{sec:conclusions}.

\section{Model-space BSE Hamiltonian and sum-rule methodology}
\label{sec:model}

\subsection{Exciton Hamiltonian}

We consider a one-dimensional lattice of $L$ sites with periodic boundary
conditions. The exciton is described by one electron in a conduction band
and one hole in a valence band. In the basis $|i,j\rangle = c_i^\dagger
h_j^\dagger |0\rangle$, the model-space BSE Hamiltonian is
\begin{widetext}
\begin{equation}
\hat{H}_X = -t_e \sum_i \left(c_i^\dagger c_{i+1} + \text{h.c.}\right)
           -t_h \sum_i \left(h_i^\dagger h_{i+1} + \text{h.c.}\right)
           + \sum_{ij} V_{eh}(i-j)\, n_i^e\, n_j^h,
\label{eq:HX}
\end{equation}
\end{widetext}
where $t_e$ and $t_h$ are the electron and hole hopping amplitudes,
$n_i^e = c_i^\dagger c_i$, $n_j^h = h_j^\dagger h_j$, and $V_{eh}(r)$
is the attractive electron-hole interaction. We adopt a soft-Coulomb form
\begin{equation}
V_{eh}(r) = -\frac{U_X}{\sqrt{r_{\min}^2 + a_c^2}},
\label{eq:Veh}
\end{equation}
where $r_{\min} = \min(|r|, L-|r|)$ is the minimum-image distance and
$a_c$ is a core radius that regularizes the on-site interaction. The
parameter $U_X$ controls the exciton binding energy and, consequently,
the localization length $\xi$: small $U_X/t_e$ yields a delocalized
Wannier exciton, while large $U_X/t_e$ yields a compact Frenkel exciton.

\subsection{Center-of-mass decomposition}
Exploiting translational invariance, we decompose the Hilbert space into blocks of definite total momentum $K$. In the relative-coordinate basis $|K, r\rangle$, where $r$ denotes the electron--hole separation, the Hamiltonian block reads
\begin{align}
H_K(r, r \pm 1) &= -t_e e^{\pm i\beta_h K} - t_h e^{\mp i\beta_e K}, \\
H_K(r, r) &= V_{eh}(r),
\end{align}
where the mass fractions are defined by the continuum center-of-mass transformation $r_e = R + \beta_h r$ and $r_h = R - \beta_e r$. Since the effective mass in a tight-binding model is inversely proportional to the hopping amplitude ($m^* \propto 1/t$), the mass fractions are $\beta_e = m_e/M = t_h/(t_e + t_h)$ and $\beta_h = m_h/M = t_e/(t_e + t_h)$. The ground-state exciton at $K=0$ has a real, even envelope $\phi_0(r) = \phi_0(-r)$, which is the origin of the symmetry protection.

\subsection{Exciton-phonon vertices}

We introduce two phonon coupling channels. The long-range
Fr\"ohlich-type interaction couples to the charge-density fluctuation:
\begin{equation}
\hat{H}_{e\text{-}ph}^{(F)} = \sum_q M_F(q)\,
\hat{\rho}_X(q)\left(b_q + b_{-q}^\dagger\right),
\label{eq:HF}
\end{equation}
where $\hat{\rho}_X(q) = \sum_j e^{iqj}(n_j^e - n_j^h)$ is the exciton
charge-density operator and $M_F(q) = g_F / \sqrt{|q| + q_s}$ is the
Fr\"ohlich kernel with infrared regulator $q_s$.

The short-range Holstein-type interaction couples to the total
exciton density:
\begin{equation}
\hat{H}_{e\text{-}ph}^{(H)} = \sum_q M_H(q)\,
\hat{D}_X(q)\left(b_q + b_{-q}^\dagger\right),
\label{eq:HH}
\end{equation}
where $\hat{D}_X(q) = \sum_j e^{iqj}(n_j^e + n_j^h)$ and
$M_H(q) = g_H$ is momentum-independent.

\subsection{Exact sum-rule evaluation}

The key methodological advance of this work is the evaluation of the
total coupling weight via the completeness relation. For an operator
$\hat{O}$, the sum over all final excitonic states satisfies
\begin{equation}
\sum_n \left|\langle n, K+q | \hat{O}(q) | 0, K \rangle\right|^2
= \langle 0, K | \hat{O}^\dagger(q)\,\hat{O}(q) | 0, K \rangle.
\label{eq:sumrule}
\end{equation}
This identity eliminates the need to diagonalize or store excited states.
For the Fr\"ohlich channel, the sum-rule weight is
\begin{widetext}
\begin{equation}
W_F(q) = \sum_r |\phi_0(r)|^2 \left|e^{iq\alpha_h r} - e^{-iq\alpha_e r}\right|^2
       = \sum_r |\phi_0(r)|^2 \left[2 - 2\cos(qr)\right],
\label{eq:WF}
\end{equation}
\end{widetext}
and for the Holstein channel,
\begin{widetext}
\begin{equation}
W_H(q) = \sum_r |\phi_0(r)|^2 \left|e^{iq\alpha_h r} + e^{-iq\alpha_e r}\right|^2
       = \sum_r |\phi_0(r)|^2 \left[2 + 2\cos(qr)\right].
\label{eq:WH}
\end{equation}
\end{widetext}
Remarkably, both expressions depend \emph{only} on the ground-state
probability density $|\phi_0(r)|^2$ and are independent of mass
asymmetry. This follows directly from the completeness of the excitonic eigenstates, 
$\sum_n |n\rangle\langle n| = \hat{I}$, and is therefore exact within the model space.

The total coupling weights are obtained by integrating over phonon
momentum:
\begin{widetext}
\begin{equation}
\mathcal{W}_F = \frac{1}{L}\sum_q |M_F(q)|^2\, W_F(q),
\qquad
\mathcal{W}_H = \frac{1}{L}\sum_q |M_H(q)|^2\, W_H(q).
\label{eq:totalW}
\end{equation}
\end{widetext}

\subsection{Intraband vertices and symmetry breaking}
While the inclusive weights [Eq.~(7)] govern high-energy Franck--Condon shifts, the \emph{intraband} channel (ground state to ground state with momentum transfer $K \to K+q$) governs low-energy decoherence, momentum scattering, and zero-phonon-line broadening. The intraband Fr\"ohlich vertex is
\begin{equation}
F_{00}(q) = \sum_r \phi_q^*(r)\phi_0(r)\left(e^{iq\beta_h r} - e^{-iq\beta_e r}\right),
\end{equation}
where $\phi_q(r)$ is the ground-state eigenvector of $H_{K=q}$.

For symmetric masses ($t_e = t_h$, hence $\beta_e = \beta_h = 1/2$), the hopping is purely real for all $K$, so $\phi_q(r)$ is real and even. The vertex $e^{iqr/2} - e^{-iqr/2} = 2i\sin(qr/2)$ is purely imaginary and odd. The integrand is therefore an odd function of $r$, and the intraband vertex vanishes exactly:
\begin{equation}
F_{00}(q) = 0 \quad \text{for} \quad m_e = m_h, \; \forall q.
\end{equation}
This is the exact symmetry protection theorem. For asymmetric masses ($t_e \neq t_h$), the hopping becomes complex, and $\phi_q(r)$ acquires a non-trivial phase twist. The integrand is no longer purely odd, and the intraband vertex becomes finite, breaking the protection.

\section{Analytical long-wavelength limit and symmetry breaking}
\label{sec:analytical}

\subsection{Gauge transformation and the phase twist}
We remain strictly within the one-dimensional model. The relative-coordinate Hamiltonian $H_K$ has off-diagonal hopping $J_K = t_e e^{i\beta_h K} + t_h e^{-i\beta_e K}$. We write $J_K = |J_K| e^{i\theta_K}$, where the phase $\theta_K$ is given by
\begin{equation}
\tan \theta_K = \frac{t_e \sin(\beta_h K) - t_h \sin(\beta_e K)}{t_e \cos(\beta_h K) + t_h \cos(\beta_e K)}.
\end{equation}
For small $K$, $\theta_K \approx K \frac{t_e \beta_h - t_h \beta_e}{t_e + t_h}$. To render the bulk Hamiltonian purely real, we apply the gauge transformation $\phi_K(r) = e^{-i\theta_K r} u_K(r)$. Since the potential $V_{eh}(r)$ is symmetric, the envelope $u_K(r)$ can be chosen as strictly real and even in $r$.

\subsection{Small-$q$ activation law}
The intraband Fr\"ohlich vertex is
\begin{equation}
F_{00}(q) = \sum_r \phi_q^*(r) \phi_0(r) \left[ e^{iq\beta_h r} - e^{-iq\beta_e r} \right].
\end{equation}
Since $\theta_0 = 0$, $\phi_0(r) = u_0(r)$. For the final state, $\phi_q^*(r) = e^{i\theta_q r} u_q(r)$. Expanding the phase factor and the vertex operator for small $q$ (the long-wavelength limit relevant to macroscopic polar fields):
\begin{align}
e^{i\theta_q r} &\approx 1 + i\theta_q r, \\
V_q(r) = e^{iq\beta_h r} - e^{-iq\beta_e r} &\approx iqr - \frac{q^2}{2}(\beta_h^2 - \beta_e^2)r^2.
\end{align}
Multiplying these and keeping terms up to $\mathcal{O}(q^2)$ that are even in $r$ (since $u_q u_0$ is even), we obtain the leading non-vanishing term:

\begin{equation}
F_{00}(q) \approx -q^2 \left( \frac{t_e \beta_h - t_h \beta_e}{t_e + t_h} + \frac{\beta_h^2 - \beta_e^2}{2} \right) \langle r^2 \rangle_{0,q}.
\end{equation}
Substituting the definitions of the mass fractions ($\beta_e = t_h/(t_e+t_h)$ and $\beta_h = t_e/(t_e+t_h)$), the coefficient simplifies exactly to:
\begin{equation}
\Delta = \frac{t_e - t_h}{t_e + t_h} + \frac{t_e - t_h}{2(t_e + t_h)} = \frac{3}{2} \frac{t_e - t_h}{t_e + t_h}.
\end{equation}
Defining this simplified asymmetry parameter as $\Delta$, we arrive at the central analytical result:
\begin{equation}
F_{00}(q) \approx -\Delta q^2 \langle r^2 \rangle.
\label{eq:Fel_final}
\end{equation}
To leading order in $q\xi$, this proves that mass asymmetry ($t_e \neq t_h$) activates a previously forbidden low-$q$ scattering channel. We emphasize that Eq.~\eqref{eq:Fel_final} is a robust long-wavelength theorem. The fully integrated weight over the Brillouin zone is sensitive to the UV cutoff where $q\xi \sim 1$, and thus does not universally inherit the simple $q^2$ scaling.

\section{Numerical results across the Wannier-Frenkel crossover}
\label{sec:results}

We now present the numerical validation of the analytical framework
developed in Sec.~\ref{sec:analytical}. All calculations are performed
on a one-dimensional ring of $L = 128$ sites with hopping amplitudes
$t_e = 1$ (setting the energy scale) and $t_h$ varied to control mass
asymmetry. The electron-hole interaction strength $U_X$ is swept from
$0.2\,t_e$ (deep Wannier regime) to $12\,t_e$ (Frenkel limit). The
Fr\"ohlich kernel uses $g_F = 1$ and infrared regulator $q_s = 0.05$;
the Holstein kernel uses $g_H = 1$.

\subsection{Exciton localization and the Wannier-Frenkel crossover}

To rigorously quantify the Wannier-Frenkel crossover, we track two complementary measures of the ground-state envelope $\phi_0(r)$. First, we define the participation ratio (PR) as
\begin{equation}
{\rm PR} = \frac{1}{\sum_r |\phi_0(r)|^4},\label{eqn:partratio}
\end{equation}
which estimates the effective number of lattice sites over which the exciton is spread. It ranges from ${\rm PR} \approx 1$ for a strictly single-site Frenkel exciton to ${\rm PR} \approx L$ for a uniformly delocalized state. Second, we define the localization length $\xi$ via the second moment of the probability density,
\begin{equation}
\xi = \sqrt{\sum_r d(r)^2 |\phi_0(r)|^2},
\end{equation}
where $d(r) = \min(|r|, L-|r|)$ is the minimum-image distance on the periodic ring. 

\begin{figure}[t]
\centering
\includegraphics[width=\columnwidth]{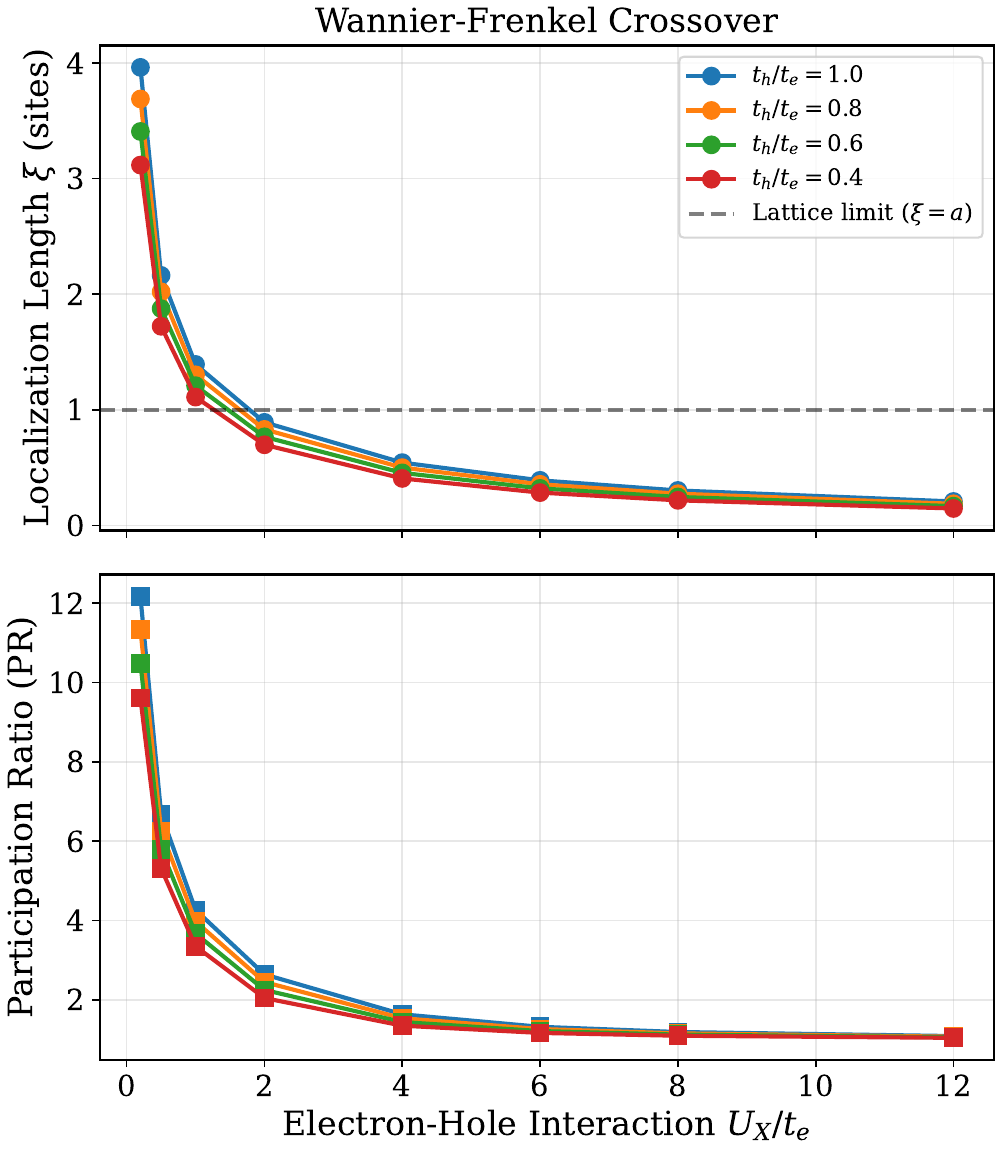}
\caption{Exciton localization length $\xi$ (top) and 
participation ratio (bottom) as functions of the electron-hole
interaction strength $U_X/t_e$ for mass ratios $t_h/t_e = 1.0$ (black),
$0.8$ (red), $0.6$ (blue), and $0.4$ (green). The horizontal dashed
line marks $\xi = 1$ (single-site Frenkel limit). The crossover from
Wannier to Frenkel character occurs near $U_X/t_e \sim 2$-$4$.}
\label{fig:localization}
\end{figure}

Figure~\ref{fig:localization} displays $\xi$ and PR as functions of the interaction strength $U_X/t_e$ for four mass ratios $t_h/t_e \in \{1.0,\, 0.8,\, 0.6,\, 0.4\}$. In the weak-coupling regime ($U_X/t_e \lesssim 0.5$), the exciton is relatively delocalized, extending over $\xi \sim 3$-$4$ lattice spacings with ${\rm PR} \sim 10$-$12$. This is characteristic of a Wannier-Mott exciton~\cite{Wannier1937,Mott1938}. As $U_X$ increases, both measures decrease monotonically. In the strong-coupling limit ($U_X = 12\,t_e$), the exciton collapses toward the origin, reaching ${\rm PR} \sim 1.1$ and an RMS radius of $\xi \sim 0.15$-$0.20$ lattice spacings. This corresponds to the strict Frenkel limit~\cite{Frenkel1931}, where the probability density is overwhelmingly concentrated on a single site ($r=0$) with only a minimal exponential tail on neighboring sites. Notably, increasing mass asymmetry ($t_h/t_e < 1$) slightly reduces $\xi$ at a fixed $U_X$, reflecting the reduced kinetic energy delocalization of the heavier carrier.

\subsection{Total coupling weights: Franck-Condon channel}

\begin{figure}[t]
\centering
\includegraphics[width=\columnwidth]{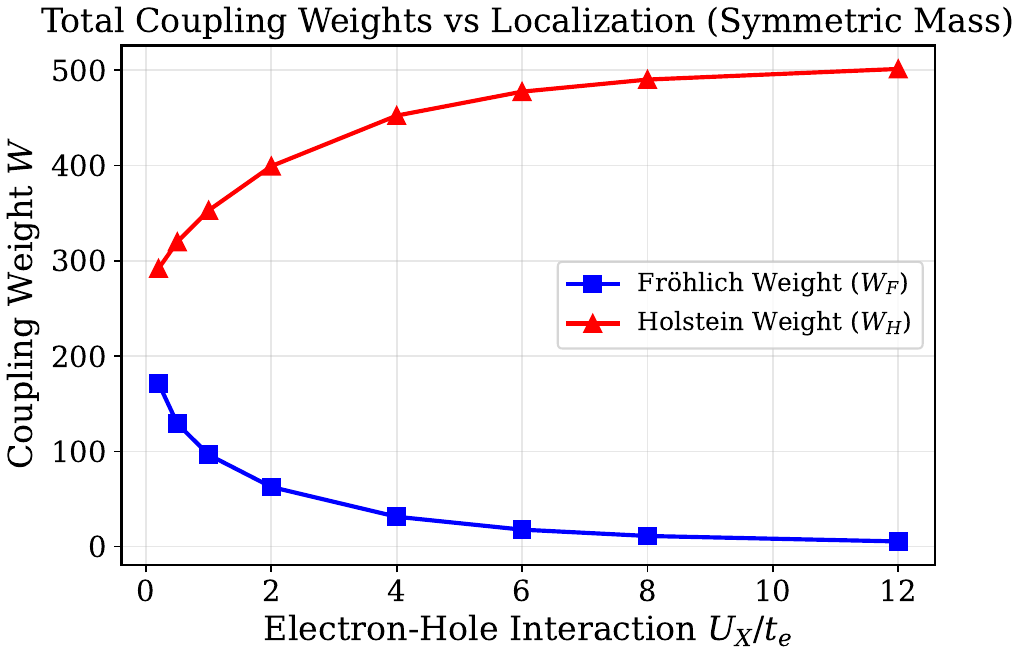}
\caption{Total Fr\"ohlich weight $\mathcal{W}_F$ (blue circles) and
Holstein weight $\mathcal{W}_H$ (red squares) versus $U_X/t_e$ for
symmetric masses ($t_h/t_e = 1$). The Fr\"ohlich weight vanishes in the
Frenkel limit while the Holstein weight saturates, yielding a crossover
ratio exceeding two orders of magnitude.}
\label{fig:totalweights}
\end{figure}

Figure~\ref{fig:totalweights} shows the total Fr\"ohlich and Holstein
coupling weights $\mathcal{W}_F$ and $\mathcal{W}_H$ [Eq.~\eqref{eq:totalW}]
as functions of $U_X$. Two key features emerge:

\begin{enumerate}
\item The total Fr\"ohlich weight $\mathcal{W}_F$ decreases monotonically
from $\sim 170$ (Wannier limit) to $\sim 3$ (Frenkel limit), vanishing
as the exciton localizes. This is the sum-rule prediction of
Eq.~\eqref{eq:WF}: as $|\phi_0(r)|^2 \to \delta(r)$, the factor
$2 - 2\cos(qr) \to 0$ for all $q$.

\item The total Holstein weight $\mathcal{W}_H$ increases from $\sim 290$
to $\sim 505$, reflecting the factor $2 + 2\cos(qr) \to 4$ as the
exciton collapses to a single site.
\end{enumerate}

The crossover ratio $R_{\text{tot}} = \mathcal{W}_H / \mathcal{W}_F$
spans nearly two orders of magnitude across the parameter range studied,
confirming that the Franck-Condon (absorption/emission) physics
transitions from Fr\"ohlich-dominated to Holstein-dominated as the
exciton localizes. This result is independent of mass asymmetry, as
predicted by the sum rule in Eq.~\eqref{eq:sumrule}.

\subsection{Intraband channel: symmetry protection and its breakdown}
The central result of this work is displayed in Fig.~3 and Table~I. For symmetric masses ($t_h/t_e = 1.0$), the intraband Fr\"ohlich weight is zero to machine precision ($\sim 10^{-30}$) at all values of $U_X$, confirming the exact symmetry protection theorem. For $t_h/t_e < 1.0$, the weight becomes finite, confirming that mass asymmetry breaks the protection.

\begin{figure}[ht]
\centering
\includegraphics[width=\columnwidth]{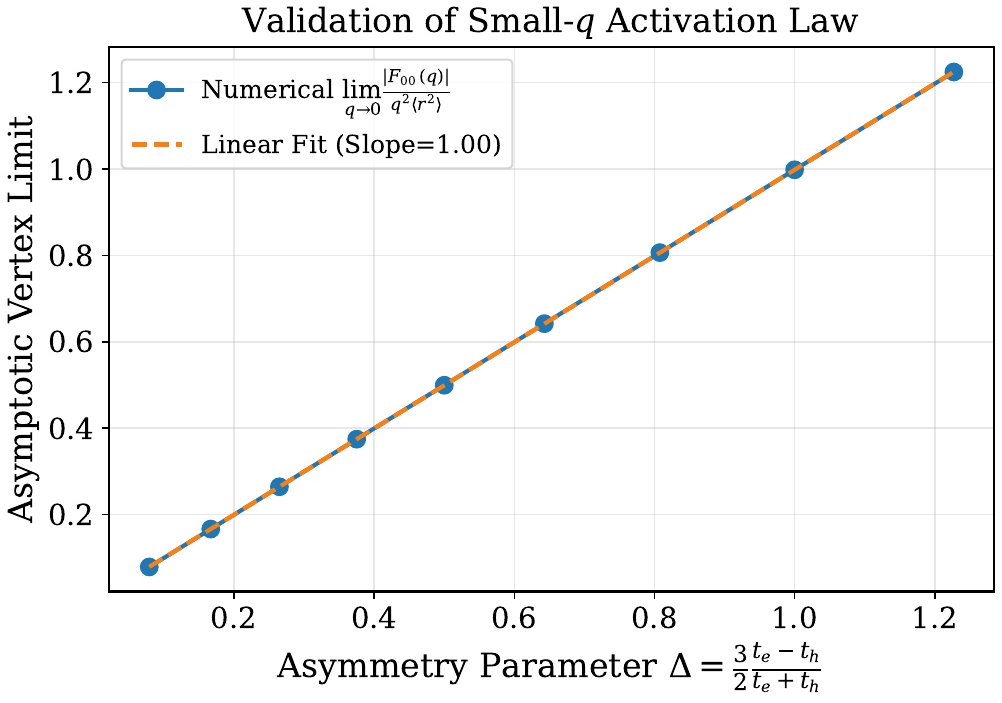}
\caption{Validation of the small-$q$ activation law. The asymptotic vertex limit $\lim_{q\to 0} |F_{00}(q)| / (q^2 \langle r^2 \rangle)$ is plotted against the simplified asymmetry parameter $\Delta = \frac{3}{2}\frac{t_e-t_h}{t_e+t_h}$ for multiple mass ratios at fixed interaction strength ($U_X = 1.5 t_e$). The data collapses onto a single straight line, confirming the exact analytical coefficient.}
\label{fig:delta}
\end{figure}

\begin{figure}[t]
\centering
\includegraphics[width=\columnwidth]{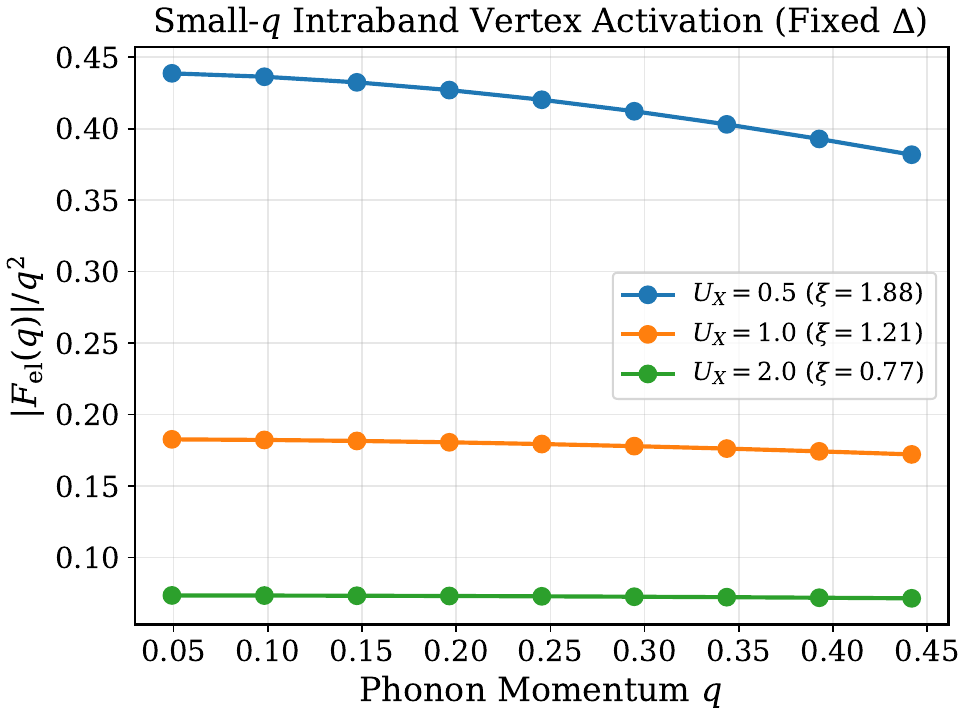}
\caption{Numerical validation of the small-$q$ activation law. The ratio $|F_{00}(q)|/q^2$ is plotted against phonon momentum $q$ for fixed mass asymmetry ($t_h/t_e = 0.6$) and varying localization lengths $\xi$. As $q \to 0$, the vertex ratio converges to a constant proportional to $\langle r^2 \rangle \sim \xi^2$, confirming Eq.~\eqref{eq:Fel_final}.}
\label{fig:small_q}
\end{figure}

To rigorously validate the small-$q$ activation law [Eq.~\eqref{eq:Fel_final}], we plot the ratio $|F_{00}(q)|/q^2$ as a function of $q$ in Fig.~\ref{fig:small_q}. For a fixed mass asymmetry ($t_h/t_e = 0.6$), the curves for different interaction strengths $U_X$ flatten out to a constant as $q \to 0$. Furthermore, the asymptotic value of this constant scales systematically with the exciton localization length $\xi$, precisely as predicted by the $\Delta q^2 \langle r^2 \rangle$ dependence. This confirms that the breakdown of symmetry protection is fundamentally a long-wavelength phenomenon. Furthermore, in Fig.~\ref{fig:delta}, we plot this asymptotic limit against the simplified asymmetry parameter $\Delta = \frac{3}{2}\frac{t_e-t_h}{t_e+t_h}$ for multiple mass ratios. The data collapses onto a single straight line, confirming the exact analytical coefficient derived in Eq.~\ref{eqn:partratio}.

\subsection{Connection to many-body observables: Decomposed self-energy}

\begin{figure}[ht]
\centering
\includegraphics[width=\columnwidth]{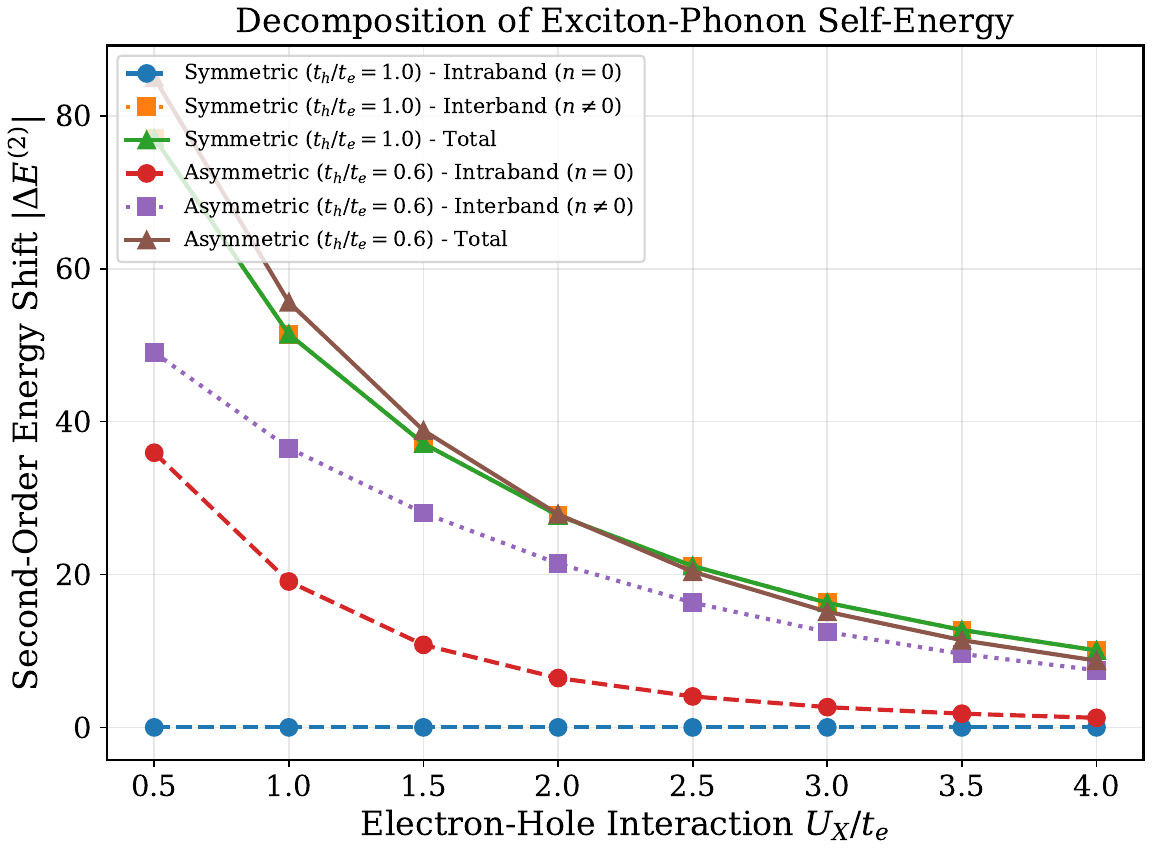}
\caption{Decomposition of the second-order exciton-phonon self-energy into intraband ($n=0$) and interband ($n \neq 0$) contributions. For symmetric masses ($t_h/t_e = 1.0$), the intraband shift is exactly zero due to symmetry protection, while the interband shift remains finite. Introducing mass asymmetry ($t_h/t_e = 0.6$) activates the intraband channel, which becomes a dominant fraction of the total energy renormalization in the intermediate localization regime.}
\label{fig:decomp}
\end{figure}

Figure~\ref{fig:decomp} displays these decomposed shifts as a function of $U_X$. For symmetric masses ($t_h/t_e = 1.0$), the intraband shift $\Delta E^{(2)}_{\text{intra}}$ is exactly zero, reflecting the symmetry protection theorem. However, the interband shift $\Delta E^{(2)}_{\text{inter}}$ remains finite, demonstrating that the exciton still couples to phonons via internal transitions.

When mass asymmetry is introduced ($t_h/t_e = 0.6$), the intraband channel is activated. Figure~\ref{fig:decomp} reveals that this newly activated $\Delta E^{(2)}_{\text{intra}}$ channel is not merely a mathematical curiosity; it grows to become a dominant fraction of the total low-energy self-energy shift in the intermediate localization regime. This explicitly demonstrates that breaking the mass symmetry turns on a physically significant low-energy scattering channel that profoundly modifies the second-order exciton energy renormalization.

\section{Discussion: Conceptual implications for excitonic materials}
\label{sec:discussion}

While the 1D tight-binding model cannot quantitatively predict material-specific linewidths, the separation of inclusive and intraband coupling, alongside the small-$q$ activation law, provides a robust conceptual framework for interpreting exciton-phonon physics in complex materials.

\subsection{Implications for polar and low-dimensional semiconductors}
In lead halide perovskites, excitons are relatively delocalized ($\xi \gg a$) and mass asymmetry is moderate. Our framework suggests that while the inclusive Franck--Condon coupling remains finite, the intraband decoherence channel is heavily suppressed by the near-symmetric masses, consistent with the weak phonon dressing and narrow linewidths observed at low temperatures. 

In monolayer transition-metal dichalcogenides (TMDs), excitons are highly confined ($\xi \sim 2a$) and exhibit strong band-structure asymmetry. The breakdown of symmetry protection via the phase twist mechanism suggests a possible mechanism that may be relevant to the strong, complex phonon sidebands observed in TMD photoluminescence. Similarly, in wide-gap transition-metal oxides like TiO$_2$, charge-transfer excitons are highly localized. In this regime, the long-range intraband Fr\"ohlich channel is geometrically starved, and the exciton--phonon interaction is dominated by local deformation-potential coupling, suggesting a mechanism that may align with the strong coupling to local metal-oxygen stretching modes observed in resonant Raman spectroscopy.

\subsection{Limitations and extensions}
The one-dimensional geometry overestimates the role of simple parity; in 2D and 3D, angular momentum selection rules and anisotropic screening will modify the exact cancellation conditions. Furthermore, the model treats phonons as dispersionless Einstein modes. Extending this framework to include realistic phonon dispersions and non-perturbative variational polaron methods will be necessary to capture strong-coupling self-trapping phenomena.

\subsection{Dimensional generalization: angular momentum selection rules and anisotropic screening}
\label{sec:dimension}

The one-dimensional model employed here exploits a discrete left--right parity symmetry. In two and three dimensions, the analogous symmetry is spatial inversion $\mathbf{r} \to -\mathbf{r}$, and the exciton eigenstates are labeled by angular momentum quantum numbers rather than a single parity index. It is important to establish which aspects of the symmetry theorem survive this generalization and which are modified.

\paragraph{Survival of the intraband protection in centrosymmetric systems.}
In $d$ dimensions, the intraband Fr\"ohlich vertex for equal masses is
\begin{equation}
F_{00}(\mathbf{q}) = \int d^d r\; \phi_0^*(\mathbf{r})\,\phi_0(\mathbf{r})\, 2i\sin\!\left(\frac{\mathbf{q}\cdot\mathbf{r}}{2}\right),
\end{equation}
where $\phi_0$ is the exciton ground-state envelope. Since $\sin(\mathbf{q}\cdot\mathbf{r}/2)$ is odd under inversion $\mathbf{r}\to-\mathbf{r}$, while $|\phi_0(\mathbf{r})|^2$ is even for any state of definite inversion parity, the integral vanishes identically in any centrosymmetric system, regardless of dimensionality. Equivalently, expanding the plane wave in spherical harmonics,
\begin{equation}
e^{i\mathbf{q}\cdot\mathbf{r}/2} = 4\pi \sum_{\ell m} i^\ell j_\ell(qr/2)\, Y_{\ell m}(\hat{\mathbf{q}})\, Y_{\ell m}^*(\hat{\mathbf{r}}),
\end{equation}
the $s$-state ($\ell=0$) matrix element samples only the $\ell=0$ component, which is $j_0(qr/2) = \sin(qr/2)/(qr/2)$. For equal masses, the $\ell=0$ parts of the electron and hole plane waves are identical and cancel exactly in the charge-difference vertex. Thus, the intraband protection is \emph{not} an artifact of one dimensionality; it is a consequence of inversion symmetry and holds for the $s$-state exciton in any centrosymmetric material.

\paragraph{Angular momentum selection rules for interband channels.}
The dimensional generalization most significantly affects the \emph{interband} scattering channels. In 1D, excited states are simply even or odd. In 3D, the exciton states carry angular momentum $(\ell, m)$, and the Fr\"ohlich vertex connects states according to dipole-like selection rules. The charge-difference operator $e^{i\mathbf{q}\cdot\beta_h\mathbf{r}} - e^{-i\mathbf{q}\cdot\beta_e\mathbf{r}}$ is odd under inversion for equal masses, so it connects the $s$-state ($\ell=0$) only to states of odd $\ell$ (i.e., $p$, $f$, \ldots):
\begin{equation}
\langle n',\ell',m' | V_F | n, 0, 0 \rangle \neq 0 \quad \text{only if } \ell' \text{ is odd}.
\end{equation}
This means the inclusive coupling weight, which sums over all intermediate states, receives contributions from the $s\to p$ channel in 3D (and $s\to p_x,\, p_y$ in 2D). The \emph{intraband} $s\to s$ channel remains forbidden by inversion symmetry. The qualitative picture—that the inclusive weight is finite while the intraband channel is protected—therefore survives in higher dimensions, but the angular momentum decomposition of the interband weight becomes richer.

\paragraph{Mass asymmetry breaking in the continuum.}
In the continuum limit (no lattice), mass asymmetry breaks the intraband protection through a mechanism distinct from the lattice phase twist discussed in Sec.~III. For unequal masses, the $\ell=0$ components of the electron and hole plane waves are $j_0(q\beta_h r)$ and $j_0(q\beta_e r)$, which are \emph{different}. The $s$-state matrix element becomes
\begin{equation}
F_{00}(q) = 4\pi \int_0^\infty r^2\, dr\; |\phi_0(r)|^2 \left[ j_0(q\beta_h r) - j_0(q\beta_e r) \right].
\end{equation}
Expanding $j_0(x) \approx 1 - x^2/6$ for small $q$, this yields $F_{00}(q) \propto (\beta_h^2 - \beta_e^2)\, q^2 \langle r^2 \rangle$, recovering the same $q^2 \langle r^2 \rangle$ scaling as the 1D model. The lattice phase-twist mechanism (Sec.~III) provides an \emph{additional} breaking channel that is absent in the continuum but does not alter the leading-order scaling. The small-$q$ activation law is therefore robust across dimensionalities and model types.

\paragraph{Non-centrosymmetric materials.}
A critical caveat for real materials is that many systems of interest \emph{lack inversion symmetry}. Monolayer transition-metal dichalcogenides (e.g., MoS$_2$, WS$_2$) are non-centrosymmetric due to the broken out-of-plane mirror symmetry. In such materials, the exciton states do not carry a definite inversion parity, and the protection theorem does not apply even for equal electron and hole masses. The intraband Fr\"ohlich channel is generically active in non-centrosymmetric 2D semiconductors, independent of mass asymmetry. This represents a fundamental distinction between centrosymmetric bulk materials (where the protection can hold) and non-centrosymmetric monolayers (where it is broken by crystal symmetry alone). Bulk TMDs (2H phase), which are centrosymmetric, would retain the protection for equal masses.

\paragraph{Anisotropic and dimensionally modified screening.}
The Fr\"ohlich kernel $M_F(q)$ encodes the dielectric response of the medium. In anisotropic three-dimensional crystals (e.g., rutile TiO$_2$, where $\epsilon_\perp \neq \epsilon_\parallel$), the kernel becomes direction-dependent:
\begin{equation}
M_F(\mathbf{q}) \propto \frac{1}{\sqrt{\mathbf{q}\cdot \boldsymbol{\epsilon}\cdot \mathbf{q}}},
\end{equation}
where $\boldsymbol{\epsilon}$ is the dielectric tensor. This modifies the \emph{weighting} of different $\mathbf{q}$-directions in the integrated self-energy but does not alter the vertex selection rule, which is a property of the exciton wavefunction and the charge-density operator, not of the screening kernel. In two-dimensional materials, the Coulomb interaction is described by the Keldysh potential rather than $1/r$, producing non-hydrogenic exciton envelopes with modified $\langle r^2 \rangle$. This changes the quantitative value of the activation coefficient $\Delta \langle r^2 \rangle$ but preserves the symmetry structure of the vertex.

\paragraph{Summary of dimensional effects.}
Table~\ref{tab:dimension} summarizes how each ingredient of the symmetry theorem is modified by dimensionality and material symmetry.

\begin{table*}[h]
\caption{Modification of the symmetry protection theorem by dimensionality, crystal symmetry, and screening. The intraband $s$-state protection is robust in any centrosymmetric system regardless of dimension. It is broken by mass asymmetry, non-centrosymmetric crystal structure, or strong spin--orbit mixing.}
\label{tab:dimension}
\begin{ruledtabular}
\begin{tabular}{lll}
Condition & Intraband protection & Mechanism \\
\hline
Centrosymmetric, $m_e = m_h$ & Protected ($F_{00}=0$) & Inversion symmetry \\
Centrosymmetric, $m_e \neq m_h$ & Broken & Bessel non-cancellation / phase twist \\
Non-centrosymmetric (e.g., monolayer TMD) & Broken & No definite parity \\
Strong spin--orbit mixing & Broken & Mixed exciton character \\
Anisotropic screening ($\epsilon_\perp \neq \epsilon_\parallel$) & Protected* & Kernel modified, vertex unchanged \\
2D Keldysh screening & Protected* & $\langle r^2 \rangle$ modified, symmetry intact \\
\end{tabular}
\end{ruledtabular}
\medskip
\small{*Protection holds provided the crystal is centrosymmetric and $m_e = m_h$.}
\end{table*}

\section{Conclusions}\label{sec:conclusions}
We have presented a model-space Bethe--Salpeter framework to evaluate exciton--phonon coupling across the extended-to-localized crossover. By employing exact completeness sum rules, we decoupled the inclusive high-energy Franck--Condon weights from the exclusive intraband scattering amplitudes. 

We proved an exact symmetry theorem: for an inversion-symmetric Hamiltonian with equal electron and hole masses, the intraband Fr\"ohlich vertex vanishes identically, protecting the exciton from low-energy polar phonon scattering. We derived a controlled small-$q$ activation law showing that mass asymmetry breaks this protection by inducing a complex phase twist in the moving exciton wavefunction, activating a scattering channel that scales as $F_{00}(q) \propto \Delta q^2 \langle r^2 \rangle$. Finally, by decomposing the second-order exciton--phonon self-energy, we demonstrated that this symmetry breaking activates a physically dominant intraband scattering channel that profoundly modifies the many-body energy renormalization of the exciton. These results provide a rigorous, conceptually unified framework for understanding the competition between long-range and local exciton--phonon coupling beyond standard phenomenological models.

\begin{acknowledgments}
The authors acknowledge the APhRICA training and collaboration program.
\end{acknowledgments}


\end{document}